\documentclass[journal]{IEEEtran}
\usepackage{cite}     
\usepackage{graphicx}

\usepackage{amsmath}

\begin{document}
\pagestyle{empty}
\title{Direct Coupling of SiPMs to Scintillator Tiles for Imaging Calorimetry and Triggering}

\author{{Frank Simon, Christian Soldner \\ {\small{Max-Planck-Institut f\"ur Physik, Munich, Germany and Excellence Cluster Universe, Technical University Munich, Germany }}\vspace{2mm}
\\Christian Joram \\{\small{CERN, Geneva, Switzerland}}

}
}

\maketitle

\begin{abstract}
The recent availability of blue sensitive silicon photomultipliers allows the direct readout of blue emitting plastic scintillator tiles without the use of a wavelength shifting fiber. Such directly read out tiles, without light guides, are attractive for the use in highly granular calorimeters that use large numbers of individual cells and in other applications where very compact designs are needed. However, the total signal amplitude and the uniformity of the response can be problematic in such cases.  We have developed a scanning setup to investigate the response of scintillator tiles with SiPM readout in detail. It was used to develop optimized scintillator tile geometries for highly granular hadronic calorimetry at future colliders and to investigate the feasibility of a SiPM readout for the trigger of the ATLAS ALFA luminosity detectors. We report on results obtained with specialized scintillator tile geometries, discuss first results obtained with directly coupled SiPM readout of the ATLAS ALFA trigger tiles and introduce the application of fiberless readout in an experiment to study the time structure of hadronic showers.
\end{abstract}


\maketitle
\thispagestyle{empty}

\section{Introduction}

Multi-cell Geiger-mode avalanche photodiodes (G-APDs), often referred to as Silicon Photomultipiers (SiPMs)\cite{Bondarenko:2000in} have a wide range of applications in high energy physics instrumentation. They provide high photon detection efficiency and insensitivity to magnetic fields with very compact devices. A large number of such devices, approximately 8\,000, have been successfully used to read out small plastic scintillator tiles in the CALICE analog hadron calorimeter\cite{Adloff:2010hb}, a physics prototype of a highly granular calorimeter for detectors at a future Linear Collider.  Since the first generation of SiPMs used in the calorimeter prototype have their sensitivity maximum in the green spectral range while the plastic scintillator emits preferentially in the blue, a wavelength shifting fiber was embedded in each scintillator cell. The SiPM was then coupled to the WLS fiber. The recent commercial availability of SiPMs with maximum efficiency in the blue spectral range now allows the direct readout of plastic scintillators without the use of a WLS fiber. This has the advantage of simplified mechanics and of a faster response due to the elimination of the additional time constant from the fiber re-emission.

\begin{figure}
\centering
\includegraphics[width=0.45\textwidth]{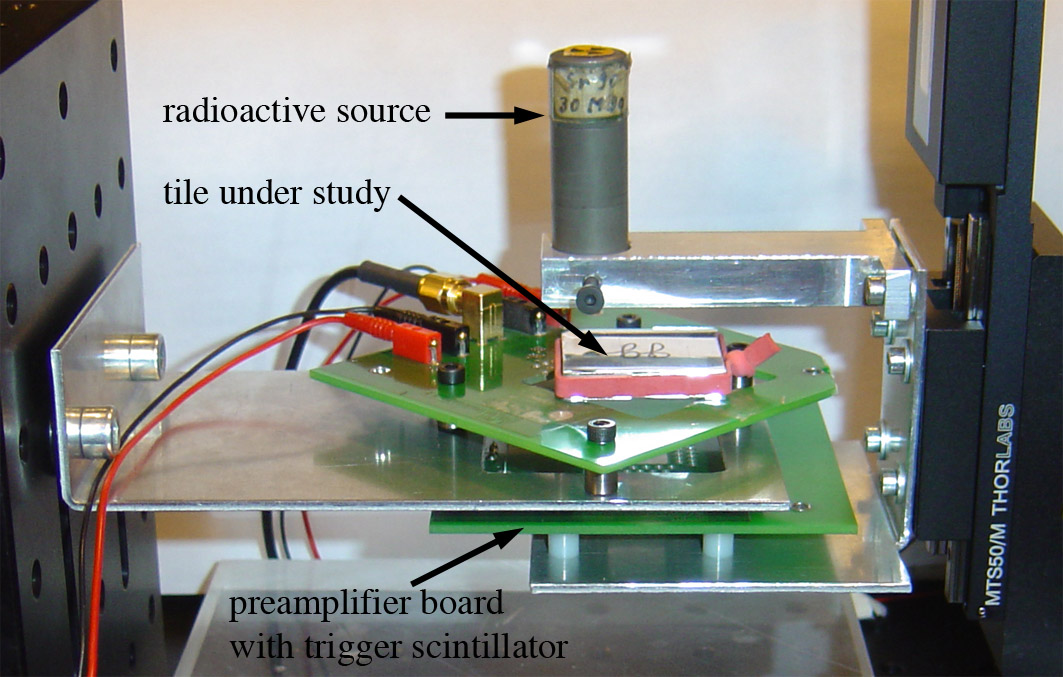}
\caption{Photograph of the scanning setup, showing the radioactive source and the scintillator under study with front end amplifier board.}
\label{fig:ScanSetup}
\end{figure}

To investigate the performance of small scintillator tiles with directly coupled SiPMs, we have constructed a scanning station using a $^{90}$Sr source mounted on a precision 3D mechanical stage \cite{Simon:2010hf}, shown in Figure \ref{fig:ScanSetup}. To ensure detection of only electrons that fully penetrate the tile under study, a small (5 mm)$^3$ trigger scintillator with SiPM readout which moves together with the source is located at the bottom of the setup.  These electrons deposit on average  about 15\% more energy than a minimum-ionizing particle \cite{Simon:2010bd} that penetrates the tile. For all studies, we use Hamamatsu MPPCs\footnote{Hamamatsu Photonics K.K., Japan. Multi Pixel Photon Counter}, with different active area and pixels size depending on the application. The photon sensors are mounted to custom made preamplifier boards and are read out with a fast oscilloscope which provides a sampling rate of 20 GS/s. This allows single photon resolution for signals of several ten photons.

\section{Uniformity of Calorimeter Cells}

Direct coupling of the photon sensor is an attractive option for highly granular calorimeters, since it offers the potential for a simplification  of fabrication and assembly, crucial for the construction of a complete detector with approximately 8 million cells with a size of $3\,\times\,3$ cm$^2$ and a thickness of 3 mm. However, a WLS fiber embedded in each scintillator cell also serves as a collector for light, leading to a good uniformity of the response over the full active area. Simple fiberless coupling of the photon sensor leads to significant non-uniformities of the response, with a strong increase of the signal close to the SiPM \cite{Simon:2010hf}. Uniformity can be reestablished by special shaping of the scintillator, with a reduction of the material close to the photon sensor by a drilled hole or depression \cite{Simon:2010hf,Blazey:2009zz}. In addition, the embedding of the SiPM into the tile leads to a significant increase in the overall signal. 

\begin{figure}
\centering
\includegraphics[width=0.495\textwidth]{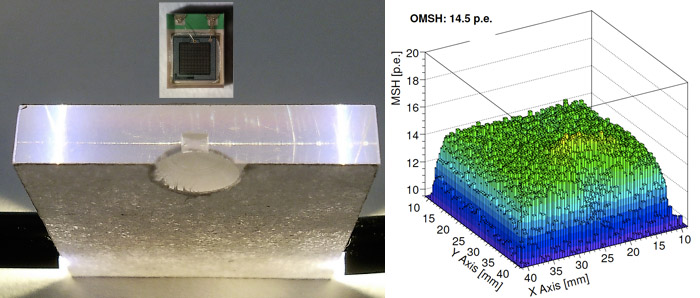}
\caption{Left: Special shaped tile of 5 mm thickness with a depression and a slit for the integration of a surface mount MPPC25-P. Right: Signal amplitude measured over the surface area, showing a high degree of uniformity. The mean amplitude is 14.5 p.e.}
\label{fig:Uniformity}
\end{figure}

Figure \ref{fig:Uniformity} shows a new tile geometry, where a spherical hole with a radius of 5 mm was drilled to a depth of 1.8 mm into
the bottom face of a 5 mm thick tile. The central axis of the sphere was 2 mm from the front edge. In addition, a slit was machined into the tile face to accommodate the photon sensor. This design is particularly well suited for mass production using molding techniques. It still needs to be adapted to a tile thickness of 3 mm, but then it is fully compatible with the readout boards used in the next generation prototype of the CALICE HCAL\cite{GoettlicherIEEE}. For the photon detection, a surface mount MPPC25-P with 1 mm$^2$ active area is used. This tile geometry provides excellent uniformity of the response as shown in the right panel of Figure \ref{fig:Uniformity}. It also yields a signal amplitude in the desired range, given by competing demands of signal to noise separation and dynamic range.

\begin{figure}
\centering
\includegraphics[width=0.495\textwidth]{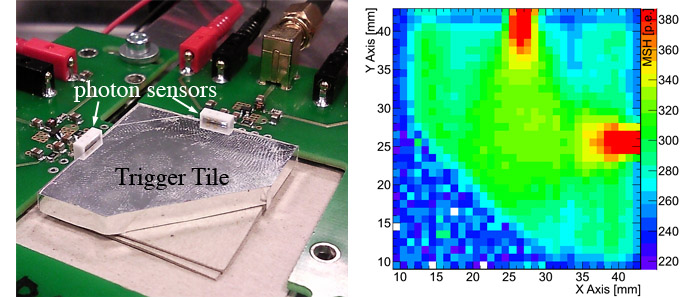}
\caption{Left: ATLAS ALFA trigger tile, 3 mm thick, coated with reflective foil, read out by two $3\,\times\,3$ mm$^2$ MPPC-50C. Right: Signal amplitude measured over the surface area, demonstrating excellent signal amplitudes over the entire tile, with a mean of 300 p.e. per penetrating electron. Due to the requirement to record 500 events at each position, the signal does not drop to zero outside of the area of the trigger tile. These signals are mostly due to scattered electrons. }
\label{fig:AlfaTile}
\end{figure}

\section{ATLAS ALFA Trigger Tiles}

Another possible application for direct readout with SiPMs are the trigger scintillators for ATLAS ALFA\cite{ATLAS_ALFA_TDR}, a scintillating fiber tracker located in Roman Pots in the LHC beam pipe on both sides of the ATLAS experiment that will provide measurements of the absolute luminosity and of low angle scattering. The main component of the ALFA detectors is a scintillating fiber tracker consisting of staggered $2\,\times\,10$ (UV) layers of fibers. The active area of the tracker is defined by a coincidence of two 3 mm thick trigger scintillators. A key requirement for the trigger is an extremely uniform efficiency over the full active area to avoid distortions of the measurement due to trigger effects. This can be best achieved by full efficiency for single particles, ensured by a high signal of the trigger detectors. In the present design, the light from the trigger scintillator is collected by bundles of clear fibers which guide it to a conventional photomultiplier located outside of the Roman Pot. Two bundles with about 100 fibers each, mounted with their front to the side faces of the trigger scintillator, are used to per trigger tile. With this setup, a satisfactory signal amplitude was achieved\cite{Braem:2009zzb}. However, the accommodation of the readout fibers of the trigger tiles  together with the fibers for the fiber tracker itself, which are also read outside the Roman Pot by MAPMTs, is challenging in the limited space available. With a readout of the trigger tiles with SiPMs directly in the Roman Pot a significantly more compact solution could be realized. The low bias voltage of a SiPM is compatible with its safe operation in the secondary vacuum of the Roman Pot.

\begin{figure}
\centering
\includegraphics[width=0.46\textwidth]{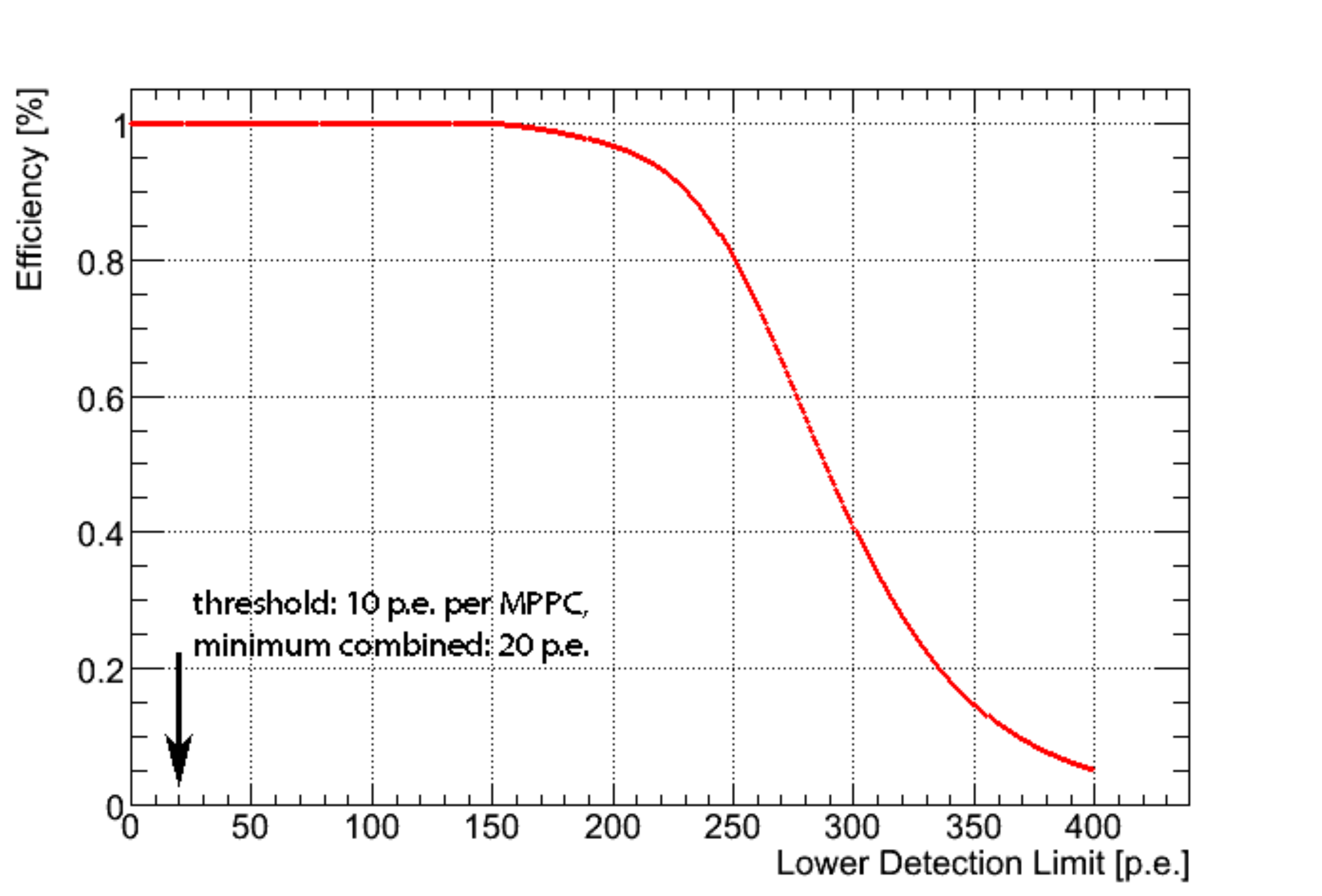}
\caption{Calculated efficiency as a function of trigger threshold on the combined signal. The arrow indicates the threshold used in the laboratory measurements.}
\label{fig:ALFAEfficiency}
\end{figure}

The ATLAS ALFA trigger tiles have a side length of 32 mm and a height of 31 mm, very close to the dimensions of the scintillator tiles of the CALICE HCAL. Thus, the same setup could be used for the response studies. The primary goal for the trigger tiles is to ensure full detection efficiency by maximizing the signal for minimum ionizing particles. To increase the number of detected photons, two photon sensors with an active area of $3\,\times\,3$ mm$^2$ and a pixel size of \mbox{$50\times 50$ $\mu$m$^2$} were used per tile, instead of the single small sensors with $25\times 25$ $\mu$m$^2$ pixels used to read out the calorimeter cells. Figure \ref{fig:AlfaTile} shows an ATLAS ALFA trigger tile with two connected MPPCs ({\it left}) and the measured amplitude distribution over the full tile area ({\it right}). The mean pulse height for a penetrating electron is around 300 photo electrons (p.e.) for both photon sensors combined, while the trigger threshold was set to 10 p.e.\ on each sensor to eliminate contributions from dark counts of the photon sensors. The signal does not fall below 250 p.e. over the full area, leading to full efficiency up to thresholds in excess of 150 p.e., thus guaranteeing uniform triggering with a comfortable margin. Figure \ref{fig:ALFAEfficiency} illustrates the efficiency for penetrating electrons from the $^{90}$Sr source. This clearly demonstrates that full efficiency for minimum ionizing particles can be achieved without problems over the full area of the trigger scintillators. The option of using SiPMs for the readout of the ATLAS ALFA trigger tiles is thus a possibility for future detector upgrades and will be investigated further. The ALFA detectors with the original fiber bundle and PMT readout will be installed during the 2010/11 winter shutdown of the LHC.

\section{Applications in Timing Studies}

One of the advantages of fiberless readout of the scintillator tiles is a faster response due to the elimination of the additional time constant from the wavelength shifter. A fast-rising signal is achieved, and the total FWHM length of the signal of minimum-ionizing particles is reduced by about a factor of two compared to scintillator tiles with embedded WLS fibers. This is used for the ``Tungsten Timing Test Beam'' (T3B) experiment, an addition to the CALICE analog hadron calorimeter to provide a first measurement of the time structure of hadronic showers in Tungsten. This experiment is motivated by the possibility of using Tungsten as absorber material for the barrel hadron calorimeter in a future CLIC detector to achieve better containment for high energy jets without having to increase the radius of the experiment's solenoid. Since timing will be a critical issue at CLIC due to the short spacing between bunch crossings, the time structure of the development of hadronic showers in a Tungsten-scintillator calorimeter is of interest. 

\begin{figure}
\centering
\includegraphics[width=0.46\textwidth]{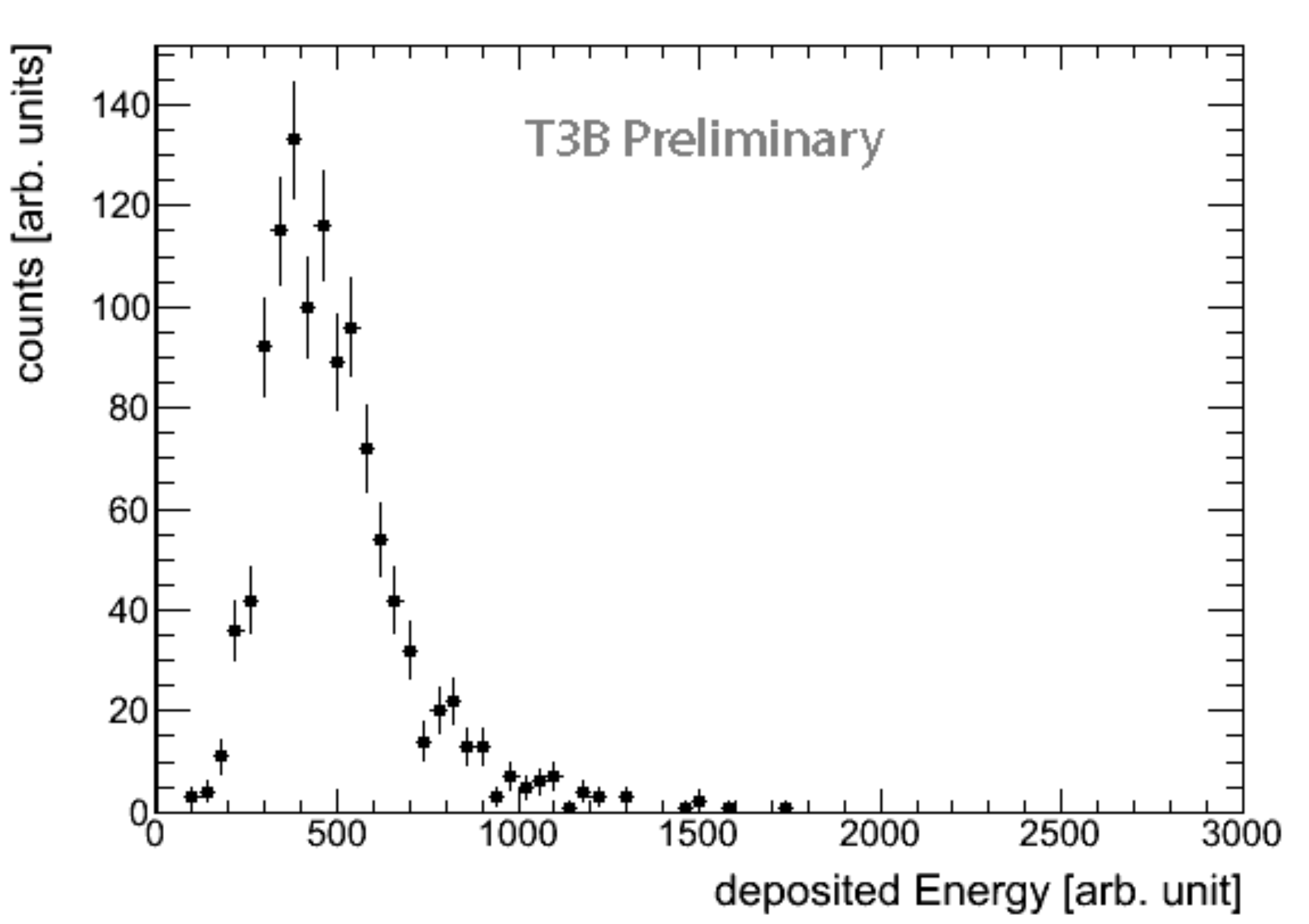}
\caption{Measured spectrum for muons in one T3B scintillator tile. The pulse height is obtained from the zero-suppressed integral of the waveform over 75 ns from the start of the pulse.}
\label{fig:T3BMIP}
\end{figure}

The T3B setup consists of 15 $3\times3$ cm$^2$ scintillator tiles with a ``dimple'' drilled into the side face at the SiPM coupling position, as discussed  above and in \cite{Simon:2010hf}. Since small signal amplitudes are important in particular for late energy deposits, the scintillators are read out with 1 mm$^2$ MPPC50P with 400 $50\times50$ $\mu$m$^2$ pixels. This results in increased photon detection efficiency compared to the measurements with MPPC25P sensors discussed above, leading to signals of approximately 26 p.e. for minimum ionizing particles. The photon sensors are read out with four-channel USB oscilloscopes\footnote{PicoTech PicoScope 6403 (http://www.picotech.com/)} with 1.25 GS per second, using long acquisition windows of 2 $\mu$s per event to record the time structure of the energy deposits in the scintillator in detail. Figure \ref{fig:T3BMIP} shows the response of one T3B scintillator tile to muons at the CERN PS measured during a commissioning campaign in September 2010. For this measurement, the T3B setup was triggered through the CALICE DAQ using external large area scintillators. By arranging three of the T3B scintillator cells as a beam telescope, the time resolution of the system was studied. In a very simple preliminary analysis using constant thresholds for all channels, a resolution of 800 ps for minimum ionizing particles was obtained. It is expected that this could be improved with more advanced signal processing techniques.

For the study of the time structure of hadronic showers in tungsten, the 15 scintillator cells are arranged in one row in the last layer (layer 31) of a prototype Tungsten calorimeter, which uses 10 mm Tungsten absorber plates and the active layers of the CALICE analog hadron calorimeter \cite{Adloff:2010hb}. The T3B scintillators are read out synchronously with the CALICE detectors, allowing the determination of the shower start point, and with that the depth of the timing measurement within the shower. Over many events, averaged time profiles can thus be measured. A first beam test with hadrons is being performed in November 2010 at the CERN PS with energies of up to \mbox{10 GeV}.

\section{Conclusions}

Plastic scintillators with Silicon Photomultiplier readout have many possible applications, including highly granular ``imaging'' calorimeters and trigger systems. A direct coupling without the use of wavelength shifters is possible for blue sensitive devices, providing fast response and mechanical simplicity. To achieve a satisfactory uniformity for hadron calorimeter applications at future colliders, a special shaping of the coupling position is required. New tile geometries, well suited for mass production using molding techniques, have shown good overall performance. The developed geometries are compatible with the next generation CALICE analog HCAL electronics. For the trigger counters of ATLAS ALFA, full efficiency can be reached by reading each scintillator tile with two photon sensor with an active area of $3\times 3$ mm$^2$ each, making the SiPM readout a viable upgrade option. Fiberlessly coupled scintillator tiles are now also used for measurements of the time structure of hadronic showers, and have already demonstrated good time resolution for single Muons.

\bibliographystyle{IEEEtran.bst}
\bibliography{CALICE}

\end{document}